\documentclass[prl,twocolumn,super,showpacs]{revtex4}
\usepackage{graphicx}
\usepackage{color}
\usepackage{amsmath}
\usepackage{latexsym}
\usepackage{amssymb}
\usepackage{bm}

\begin{document}
\title{Collective Thermotaxis of Thermally Active Colloids}

\author{Ramin Golestanian}
\email[]{ramin.golestanian@physics.ox.ac.uk}
\affiliation{Rudolf Peierls Centre for Theoretical Physics, University of Oxford, Oxford OX1 3NP, UK}

\date{\today}

\begin{abstract}
Colloids with patchy metal coating under laser irradiation could act as local sources of heat
due to the absorption of light. While for asymmetric colloids this could induce self-propulsion,
it also leads to the generation of a slowly decaying temperature profile that
other colloids could interact with. The collective behavior of a dilute solution of such
thermally active particles is studied using a stochastic formulation. It is found that when the Soret
coefficient is positive, the system could be described in stationary-state by the nonlinear
Poisson-Boltzmann equation and could adopt density profiles with significant depletion in
the middle region when confined. For colloids with negative Soret coefficient, the system
can be described as a dissipative equivalent of a gravitational system. It is shown that
in this case the thermally active colloidal solution could undergo an instability
at a critical laser intensity, which has similarities to supernova explosion.
\end{abstract}

\pacs{82.70.Dd,05.70.Ln,47.57.-s,47.70.-n}
\maketitle

The motion of colloidal particles in a solution in the presence of an externally 
applied temperature gradient, which is known as thermophoresis or the Soret effect \cite{Ludwig}, 
has been studied since the $19^{\rm th}$ century and observed in a variety of systems
\cite{Soret-observe}. Although its existence can be well formulated in 
non-equilibrium thermodynamics within linear response theory \cite{de Groot}, 
many aspects of the microscopic nature of the phenomenon has remained a subject of 
investigation to date \cite{Soret-mechanism}. The effect has also been shown to provide
a powerful tool for manipulating macromolecules and colloids \cite{Soret-manipulate}. 
Since in the phoretic transport mechanisms the colloids experience no net force,
it is possible to take advantage of them to design self-propelled particles 
by incorporating a built-in mechanism that provides asymmetric sources 
that could generate and maintain the necessary gradient across them needed for 
propulsion \cite{self-diff,gla-2}. Recently, Jiang et al. \cite{Sano-1} have 
shown that silica beads half-coated with gold when irradiated with a defocused 
laser beam exhibit such a propulsion, as the gold caps act as heat sources 
when they absorb light. Moreover, even without the self-propulsion, laser-heated 
gold-coated colloids have been shown to undergo substantially enhanced Brownian 
diffusion, which is related to the modification of the temperature in 
the medium and the resulting changes in the viscosity \cite{hot}. Since such 
thermally active colloids would create temperature profiles around them that 
decay as $1/r$, in addition to causing them to self-propel thermophoresis 
could provide a mechanism for them to interact with one another in a solution. 
The long-ranged nature of the inter-colloidal thermophoretic interaction 
could lead to interesting collective behaviors.

\begin{figure}[b]
\includegraphics[width=0.80\linewidth]{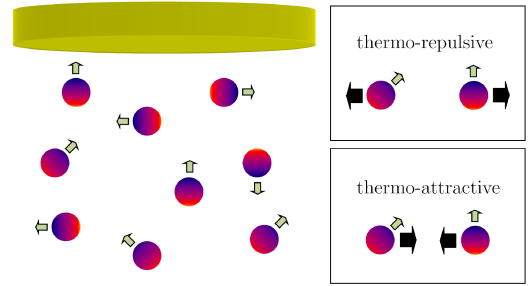}
\caption{(color online.) The metal-coated Janus-spheres under irradiation could self-propel in 
the directions shown by the (green) arrows and interact with one-another via 
the long-ranged temperature profiles they generate. The interactions are mutually 
repulsive when $S_T>0$ and attractive when $S_T<0$.
\label{fig:schem}}
\end{figure}

Here we construct a stochastic formulation to describe the collective behavior of
a number of thermally active colloids. At the long time and large length scale
limit and for dilute solutions, the formulation simplifies to a set of two nonlinear
coupled differential equations for the density and temperature profiles in the medium.
In stationary-state, we provide a number of examples for which the equations could be
solved exactly. They show a depletion effect for the case of positive Soret coefficient, 
and an instability at a finite laser intensity for negative Soret coefficient.

We consider $N$ colloidal particles of radius $R$ that are half-coated with a metal that
absorbs the laser light with an efficiency $\epsilon$, thus creating a local source of heat
of magnitude $\epsilon I$, where $I$ is the intensity of the laser (see Fig. \ref{fig:schem}).
We assume that the laser intensity is uniform throughout the space, and thus ignore any optical
confinement effect. In an externally generated temperature gradient, the colloids move
with a drift velocity ${\bf v}=-D_{T} \nabla T$, where $D_{T}$ is the thermodiffusion
coefficient. The asymmetric heat generation around each colloid
provides a mechanism to create and maintain a local temperature gradient that leads
to propulsion via self-thermophoresis \cite{gla-2}. For a Janus-sphere colloid,
the propulsion velocity can be calculated as $v_0=\epsilon I D_T/(6 \kappa)$,
where $\kappa$ is the thermal conductivity of the medium \cite{gla-2,Sano-1,note-1}.
The stochastic motion of the $i$-th colloid is described by its instantaneous
position ${\bf r}_i(t)$ and orientation ${\bf n}_i(t)$ that is a unit vector.
They satisfy the Langevin equations
\begin{math}
\frac{d {\bf r}_i}{dt}=v_0 {\bf n}_i-D_{T} \nabla T({\bf r}_i)+{\bm \xi}_i
\end{math}
and 
\begin{math}
\frac{d {\bf n}_i}{dt}={\bm \eta}_i \times {\bf n}_i,
\end{math}
in which ${\bm \xi}_i$ and ${\bm \eta}_i$ are Gaussian-distributed noise terms.
Since each colloid generates heat with an overall (surface-average) rate of
$\frac{1}{2} \epsilon I$, the temperature profile at the location of each
colloid is affected by the heat generated by all the other colloids. Since
the temperature profile equilibrates considerably faster than the colloids,
we have
\begin{math}
T({\bf r},t)=T_0+\frac{\epsilon I R^2}{2 \kappa} \sum_{j} \frac{1}{\left|{\bf r}-{\bf r}_j(t)\right|},
\end{math}
to the lowest order in the multipole expansion of the heat source distribution.
While this approximation should be valid for sufficiently dilute colloidal
solutions, the contributions from higher multipoles could readily be added
to the above temperature profile. To probe the colloidal activity at shorter times than 
the rotational diffusion time, we need to incorporate the time dependence of 
heat diffusion, which could lead to anomalous dynamics of the colloid \cite{msd}.

The Fokker-Planck equation for the probability distribution
${\cal P}({\bf r},{\bf n},t) \equiv \left\langle \sum_{i=1}^N
\delta\left({\bf r}-{\bf r}_i(t)\right) \delta\left({\bf n}-{\bf n}_i(t)\right)\right \rangle$,
can be constructed from the Langevin equations as
\begin{equation}
\partial_t {\cal P}+\nabla \cdot \Bigl[v_0 {\bf n} {\cal P}-D_{T} \left(\nabla T\right) {\cal P}
-D \nabla {\cal P}\Bigr]-D_r {\bf {\cal R}}^2 {\cal P}=0, \label{eq:FP}
\end{equation}
where ${\bf {\cal R}} \equiv {\bf n} \times \partial_{\bf n}$. In Eq. (\ref{eq:FP}),
$D$ and $D_r$ are the translational and rotational diffusion coefficients, respectively,
and represent the corresponding widths of the Gaussian probability distributions
for the noise terms ${\bm \xi}_i$ and ${\bm \eta}_i$ in the Langevin equations.
In a medium with uniform temperature $T$, we have $D={k_{\rm B} T}/{(6 \pi \eta R)}$
and $D_r=k_{\rm B} T/(8 \pi \eta R^3)$, where $\eta$ is the viscosity of water.
Equation (\ref{eq:FP}) should be complemented with the heat diffusion equation
\begin{equation}
-\nabla^2 T=\frac{2 \pi \epsilon I R^2}{\kappa} \int_{{\bf n}} {\cal P}({\bf r},{\bf n}),\label{eq:Poisson-1}
\end{equation}
which describes how the temperature profile is affected by the spatial distribution
of the colloids due to their role as motile heat sources. Equations (\ref{eq:FP})
and (\ref{eq:Poisson-1}) should be self-consistently solved to obtained the probability distribution
of the colloids as well as the temperature profile in the medium.

Let us define the density
$\rho({\bf r})=\int_{{\bf n}} {\cal P}({\bf r},{\bf n})$, the polarization field
${\bf p}({\bf r})=\int_{{\bf n}} {\bf n} \; {\cal P}({\bf r},{\bf n})$, and the nematic
order parameter
${\bf Q}({\bf r})=\int_{{\bf n}} \left[{\bf n} {\bf n}-\frac{1}{3} {\bf I}\right] {\cal P}({\bf r},{\bf n})$.
Performing $\int_{{\bf n}}$ on Eq. (\ref{eq:FP}), we can obtain an equation for the density as
\begin{math}
\partial_t \rho+v_0 \nabla \cdot {\bf p}-\nabla \cdot \Bigl[D_{T} \left(\nabla T\right) \rho
+D \nabla \rho \Bigr]=0,
\end{math}
which is incomplete since it has a source term in the form of $-v_0 \nabla \cdot {\bf p}$,
which is present due to the self-propulsion of the colloids. Performing $\int_{{\bf n}} {\bf n} \times$
Eq. (\ref{eq:FP}), we can obtain an equation for the polarization field as
\begin{eqnarray}
&&\partial_t {\bf p}+2 D_r {\bf p}+\frac{v_0}{3} \nabla \rho+v_0 \nabla \cdot {\bf Q}({\bf r}) \nonumber \\
&&-\nabla \cdot \Bigl[D_{T} \left(\nabla T\right) {\bf p}
+D \nabla {\bf p}\Bigr]=0, \label{eq:p-1}
\end{eqnarray}
where ${\bf {\cal R}}^2 {\bf n}=-2 {\bf n}$ is used. Equation (\ref{eq:p-1}) is also incomplete
as it depends on ${\bf Q}$, and this hierarchy will continue for higher order cumulants.

To make further progress, we can seek to truncate the hierarchy and simplify Eq. (\ref{eq:p-1})
in some approximation. At time scales much longer than $1/D_r$, the time derivative term in
Eq. (\ref{eq:p-1}) is considerably smaller than $2 D_r {\bf p}$ and can thus be ignored.
For sufficiently dilute solutions, i.e. when $\rho R^3 \ll 1$, and in the absence of any external means
that could induce polarization, such as external magnetic field for particles with a magnetic dipole
moment or gravity \cite{lyderic,holger}, we can ignore the $\nabla \cdot {\bf Q}({\bf r})$ term.
Since the time scale is much longer $1/D_r$, any transient or initial ordering would also have decayed.
Finally, at length scales much larger than $R$, the gradient terms in Eq. (\ref{eq:p-1}) can be 
neglected, and the equation can be approximated as
\begin{equation}
{\bf p} \simeq -\frac{v_0}{6 D_r} \nabla \rho.\label{eq:closing}
\end{equation}
Using Eq. (\ref{eq:closing}), we can estimate that ${\bf Q} \sim \nabla \rho \nabla \rho$, ignoring which
is consistent with our assumption of a dilute solution. Putting Eq. (\ref{eq:closing}) back
in the density equation, 
we find
\begin{equation}
\partial_t \rho-\nabla \cdot \Bigl[D_{\rm eff} \nabla \rho+D_{T} \left(\nabla T\right) \rho \Bigr]=0, \label{eq:rho-2}
\end{equation}
where $D_{\rm eff}=D+{v_0^2}/{(6 D_r)}$ is the enhanced effective diffusion
coefficient for the self-propelled active colloid \cite{self-diff,msd}, which could also be
rewritten as $D_{\rm eff}=D \left[1+\frac{2}{9} {\rm Pe}^2\right]$  (for a sphere) in terms of
the Peclet number ${\rm Pe}=v_0 R/D$ \cite{lyderic}. Equation (\ref{eq:rho-2}) should be solved in conjunction
with the heat diffusion equation, which reads $-\nabla^2 T=\frac{2 \pi \epsilon I R^2}{\kappa} \rho$.

In stationary state, Eq. (\ref{eq:rho-2}) is satisfied if
$D_{\rm eff} \nabla \rho+D_{T} \left(\nabla T\right) \rho=0$, which can be written as
$\nabla \ln \rho=-\frac{D_T}{D_{\rm eff}} \nabla T$. If we ignore the temperature dependence
in $D_{\rm eff}$, and use the Soret coefficient $S_T=D_T/D$, this can be integrated to yield
\begin{equation}
\rho({\bf r})=\rho_0 \;\exp\left\{-\frac{S_T \bigl[T({\bf r})-T_0\bigr]}
{\left(1+\frac{2}{9} {\rm Pe}^2\right)}\right\}.\label{eq:rhoT}
\end{equation}
Putting the above equation back in the heat diffusion equation yields a single nonlinear 
equation for the temperature profile as
\begin{equation}
-\nabla^2 T=\frac{2 \pi \epsilon I R^2 \rho_0}{\kappa} \;\exp\left\{-\frac{S_T \bigl[T({\bf r})-T_0\bigr]}
{\left(1+\frac{2}{9} {\rm Pe}^2\right)}\right\}.\label{eq:PB-T-1}
\end{equation}
Equation (\ref{eq:PB-T-1}), which is reminiscent of the Poisson-Boltzmann equation for electrolytes (see below),
could be solved for the temperature profile, which then yields the stationary-state density profile of
the colloids via Eq. (\ref{eq:rhoT}). Note that the Soret coefficient could be both positive and negative.

We can define an appropriate dimensionless temperature as
$\Psi \equiv \frac{|S_T| [T({\bf r})-T_0]}{1+\frac{2}{9} {\rm Pe}^2}$, and a characteristic length scale
\begin{equation}
\ell\equiv \frac{\epsilon I R^2 |S_T|}{2 \kappa\left(1+\frac{2}{9} {\rm Pe}^2\right)},\label{eq:Bjerrum}
\end{equation}
which is reminiscent of the Bjerrum length in the electrostatic analogy. We have
\begin{equation}
-\nabla^2 \Psi=k^2 e^{\mp\Psi},\label{eq:PB-T-2}
\end{equation}
where $k^2=4 \pi \ell \rho_0$, and the sign choice is $-{\rm sgn}(S_T)$. Equation (\ref{eq:PB-T-2})
is subject to the constraint $N=\rho_0 \int_{\bf r} e^{\mp\Psi}$.

We now discuss a number of interesting exact solutions for the stationary-state described by
Eq. (\ref{eq:PB-T-2}). We consider colloidal solutions that are confined in some region in space
and examine the effect of the dimensionality of the confinement geometry as well as the nature
and the strength of the thermotactic coupling. The confinement could in practice come from
the trapping effect of nonuniform laser beams, which could provide a very powerful tuning parameter.
For simplicity, we model the confinement by introducing sharp boundaries such as confining walls.

When $S_T >0$, the electrostatic analogy is complete as the colloidal particles mutually
repel one another, and the overall heat flux coming out of the solution through
the boundaries of the confining ``cage'' is reminiscent of the electric field flux lines, which 
could be thought of as an outer shell of opposite charges maintaining neutrality and stability.
We denote the $S_T >0$ case as {\em thermo-repulsive}.
This problem can be solved exactly for 1D and 2D confinements, and numerically for the 3D case \cite{PB}.
When the colloidal solution is confined between two plates of lateral size $L$ and distance $2 h$,
the density profile of the colloids is given as
\begin{equation}
\rho(x)=\frac{\rho_0}{\left[1+\frac{2 \pi^2 \ell^2}{k^2} \left(\frac{N}{L^2}\right)^2\right] \cos^2 \left(\frac{k x}{\sqrt{2}}\right)},\label{eq:sol-1D-rho-1}
\end{equation}
where $\rho_0$ is the concentration at the edge of confinement (wall), and $k$ satisfies
$\left(\frac{k h}{\sqrt{2}}\right) \tan\left(\frac{k h}{\sqrt{2}}\right)=\pi \ell h \left(\frac{N}{L^2}\right)$.
The profile of Eq. (\ref{eq:sol-1D-rho-1}) describes an accumulation of the colloids near the confining 
boundary (that is reminiscent to counterion condensation \cite{yan}) and a corresponding depletion of the central 
region of the system. In the strong coupling limit when $N \ell h/L^2 \gg 1$, we can find an approximate 
solution to the transcendental equation as
$k h \simeq \frac{\pi}{\sqrt{2}} \left[1-\frac{1}{\pi \left({N \ell h}/{L^2}\right)}\right]$.
In this limit, the ratio between the density of the colloids in the middle and at the edge
can be found as
\begin{math}
\frac{\rho_{\rm m}}{\rho_0} \simeq \frac{1}{4} \;(N \ell h/L^2)^{-2},
\end{math}
which shows a significant depletion effect. Note that the depletion becomes stronger
as $h$ is increased, when other parameters are kept fixed. The length $1/[2 \pi \ell (N/L^2)]$ 
is equivalent to the Gouy-Chapman length in the electrostatic analogy \cite{yan}. 
For a colloidal solution trapped in a cylindrical cage of length $L$ and width $2 h$, the
density profile reads
\begin{equation}
\rho(r)=\frac{\rho_0}{\left[1+\frac{1}{2}\left(\frac{N \ell}{L}\right)\right]^2
\left[1-\frac{1}{8} k^2 r^2\right]^2},\label{eq:sol-2D-rho-1}
\end{equation}
where
$k h=\sqrt{\frac{8 (N \ell/L)}{2+(N \ell/L)}}$. The strong coupling limit in this geometry
corresponds to $N \ell/L \gg 1$, in which case we have
\begin{math}
\frac{\rho_{\rm m}}{\rho_0} \simeq 4 \;(N \ell/L)^{-2}.
\end{math}
Note that the magnitude of depletion is independent of the confinement size in this geometry.
The ratio $\ell N/L$ is analogous to the so-called Manning-Oosawa parameter for highly charged 
rodlike polyelectrolytes \cite{yan}.
A similar profile can be found when the colloidal solution is confined to a spherical cage
of diameter $2 h$, where in the strong coupling limit that corresponds to $N \ell/h \gg 1$
in this case, we have
\begin{math}
\frac{\rho_{\rm m}}{\rho_0} \simeq 21.4 \;(N \ell/h)^{-2}.
\end{math}
Here, the depletion is inversely related to the size of the cage, namely it
decreases for larger confinement sizes.

\begin{figure}[t]
\includegraphics[width=0.80\linewidth]{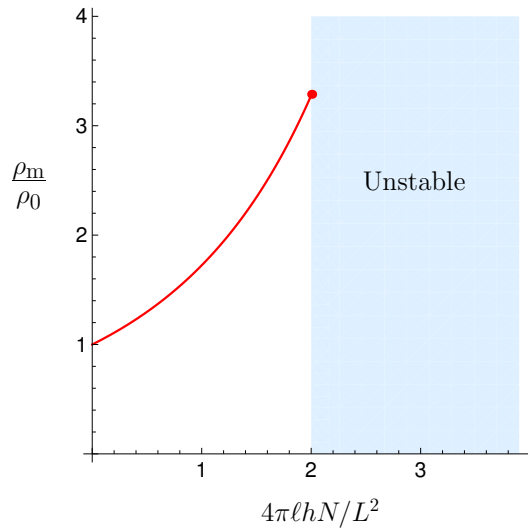}
\caption{(color online.) The density of colloids in the middle of the confined space relative to the density at the edge
as a function of the dimensionless thermophoretic coupling constant for the thermo-attractive ($S_T<0$) case.
The critical density at the onset of instability is $(\rho_{\rm m}/\rho_0)_c=3.29$, which occurs at $(4 \pi \ell h N/L^2)_c=2$.
\label{fig:inst}}
\end{figure}

When $S_T <0$, the colloids attract each other and the problem is analogous to a
gravitational system. We thus denote this case as {\em thermo-attractive}.
Let us go back to the 1D confinement geometry, where the relevant thermophoretic coupling constant
is $N \ell h/L^2$ as discussed above. In this case, Eq. (\ref{eq:PB-T-2}) (with the positive sign choice)
can be integrated in closed form and the density profile can be calculated. The stationary-state
density profile (not presented here for brevity) shows that the particles will accumulate towards
the center of the confined area. Figure \ref{fig:inst} shows that the ratio between the density in the middle
and at the edge of the confinement region increases as the thermophoretic coupling constant increases,
up to a critical point beyond which a stable (stationary-state) solution no longer exists.
The onset of instability occurs at $(4 \pi \ell h N/L^2)_c=2$, at which $(\rho_{\rm m}/\rho_0)_c=3.29$.
Similar instabilities exist in the 2D and 3D confinement cases \cite{LanLif}, where $N \ell/L$
and $N \ell/h$ play the role of the thermophoretic coupling constant, respectively.

The instability occurs because the particles that act as heat sources attract each other and
could result in a suspension that becomes increasingly denser and hotter. In this case, the heat flux at
the outer boundary of the system cannot balance the heat generated inside the confined region, which leads to
an uncontrolled buildup of thermal energy. In molecular systems, an equation of the form of
Eq. (\ref{eq:PB-T-1}) (with $S_T<0$) is used to describe exothermic combustion reactions that could
lead to thermal explosion \cite{LanLif}. It is not possible to predict what happens
in our colloidal system above the onset of this instability using the present formulation, as
the approximations used in its derivation are no longer valid. Within our handwaving analogy
to a gravitational system, this explosion would have similarities to a type I supernova
for a white dwarf, for which accreted material from the surroundings accelerates exothermic nuclear
reaction to the point that the system becomes unstable \cite{supernova}. This analogy is very rough,
however, as the colloidal system operates in the dissipative regime as opposed to the inertial
and conserved dynamics of the gravitational system.

We can estimate the length scale $\ell$ that characterizes the strength of thermophoretic interactions
from the experiment of Ref. \cite{Sano-1}. For $R=1 \;\mu$m, we can estimate that for a fully
coated bead that is not self-propelled due to lack of asymmetry (${\rm Pe}=0$),
$\ell \sim 10 \;\mu$m, while for the self-propelled colloids we could have a reduction by two orders of
magnitude, namely, $\ell \sim 0.1 \;\mu$m. Considering the confinement length to be $h \sim 10-100 \;\mu$m,
we find that it is very easy to realize a sufficiently dilute experimental system, which is in
the strong coupling limit. While the laser intensity provides a continuous tuning parameter, the presence
or absence of self-propulsion could move the system much faster in the parameter space. We note that
Eq. (\ref{eq:rho-2}) could be used to study the time dependence of the nonlinear dynamics of the colloids
as in the analogous electrokinetic system \cite{RG00}. 

The are a number of effects that we have not considered in the present analysis. We have neglected
the temperature dependence of $D_{\rm eff}$ in the calculation that led to Eq. (\ref{eq:rhoT}),
which could introduce corrections of the order of $\Delta T/T$ to the argument of the exponential.
Moreover, hydrodynamic interactions have been shown to lead to nonlocal relations
between the temperature profile and the diffusion coefficient of tracer particles \cite{RGAA}.
However, we do not expect these effects to change the qualitative behavior of the system.
We have also neglected the hydrodynamic interaction between the colloidal particles themselves.
This is justified, as self-thermophoretic colloids are effectively source dipoles and lead to
velocity fields that decay as $1/r^3$, which is faster than the thermophoretic interaction
that decays as $1/r^2$.

In conclusion, we have studied the collective behavior of active colloids that act as mobile 
heat sources, and found that thermo-repulsive colloids could organize into hollow bands, tubes, or shells, 
depending on the geometry, while thermo-attractive colloids could go unstable. We note that similar equations
could be used to study collective chemotaxis of diffusiophoretically active particles using the analogy
of nonequilibrium phoretic phenomena \cite{gla-2}.

This work was supported by EPSRC.


\begin{thebibliography}{99}

\bibitem{Ludwig}
C. Ludwig, Sitz. ber. Akad. Wiss. Wien Math.-Nat. wiss. Kl {\bf 20},
539 (1856); C. Soret, Arch. Geneve {\bf 3}, 48 (1879).

\bibitem{Soret-observe}
R. Piazza and A. Guarino, Phys. Rev. Lett. {\bf 88}, 208302 (2002); 
J. Lenglet {\em et al.}, Phys. Rev. E {\bf 65}, 031408 (2002); 
S. Duhr and D. Braun, Proc. Natl. Acad. Sci. U.S.A. {\bf 103}, 19678
(2006); C. Debuschewitz and W. K\"{o}hler, {Phys. Rev. Lett.} {\bf  87},
055901 (2001); S. Wiegand, J. Phys.: Condens. Matter {\bf 16}, R357
(2004); S.A. Putnam {\em et al.}, {Langmuir} {\bf 23}, 9221 (2007).

\bibitem{de Groot}
S.R. de Groot and P. Mazur, {\em Non-Equilibrium Thermodynamics}
(Dover, New York, 1984).

\bibitem{Soret-mechanism}
F. Brochard and P.-G. de Gennes, C.R. Acad. Sci. Paris, Serie II {\bf 293}, 1025 (1981); 
S. Fayolle {\em et al.}, Phys. Rev. Lett. {\bf 95}, 208301 (2005); 
J.K.G. Dhont {\em{et al.}}, {Langmuir} {\bf 23}, 1674 (2007);
A. W\"{u}rger, Phys. Rev. Lett. {\bf 98}, 138301 (2007); 
R.D. Astumian, Proc. Nat. Acad. Sci. USA {\bf 104}, 3 (2007);
S. N. Rasuli and R. Golestanian, Phys. Rev. Lett. 101, 108301 (2008);
A. W\"{u}rger, Phys. Rev. Lett. {\bf 101}, 108302 (2008).

\bibitem{Soret-manipulate}
D. Braun and A. Libchaber, Phys. Rev. Lett. {\bf 89}, 188103 (2002);
H.-R. Jiang {\em et al.}, Phys. Rev. Lett. {\bf 102}, 208301 (2009).

\bibitem{self-diff}
R. Golestanian, T. B. Liverpool, and A. Ajdari, Phys. Rev. Lett. {\bf 94}, 220801 (2005);
J.R. Howse {\em et al.}, Phys. Rev. Lett. {\bf 99}, 048102 (2007).

\bibitem{gla-2}
R. Golestanian, T.B. Liverpool, and A. Ajdari, New J. Phys. {\bf 9}, 126 (2007).

\bibitem{Sano-1}
H.-R. Jiang, N. Yoshinaga, and M. Sano, Phys. Rev. Lett. {\bf 105}, 268302 (2010).

\bibitem{hot}
D. Rings {\em et al.},
Phys. Rev. Lett. {\bf 105}, 090604 (2010).

\bibitem{note-1}
For simplicity, we have ignored the difference between the thermal conductivity
of water and the colloid.

\bibitem{msd}
R. Golestanian, Phys. Rev. Lett. {\bf 102}, 188305 (2009).

\bibitem{lyderic}
J. Palacci {\em et al.}, Phys. Rev. Lett. {\bf 105}, 088304 (2010).

\bibitem{holger}
M. Enculescu and H. Stark, Phys. Rev. Lett. {\bf 107}, 058301 (2011).

\bibitem{yan}
Y. Levin, Rep. Prog. Phys. {\bf 65}, 1577 (2002).

\bibitem{PB}
H.-K. Tsao, J. Phys. Chem. B {\bf 102}, 10243 (1998).

\bibitem{LanLif}
L.D. Landau and E.M. Lifshitz, {\it Fluid Mechanics} 2nd edition
(Pergamon, Oxford, England, 1987).

\bibitem{supernova}
K. Nomoto, Astrophys. J. {\bf 253}, 798 (1982).

\bibitem{RG00}
R. Golestanian, Europhys. Lett. {\bf 52}, 47 (2000).

\bibitem{RGAA}
R. Golestanian and A. Ajdari, Europhys. Lett. {\bf 59}, 800 (2002).


\end{thebibliography}
\end{document}